\documentclass[useAMS,usenatbib]{mn2e}
\usepackage{times}

\usepackage{graphicx}  
\usepackage{amsmath}
\usepackage{soul}
\usepackage{color}
\usepackage{hyperref}
\hypersetup{colorlinks,%
 citecolor=blue,%
 linkcolor=blue}
\definecolor{lightblue}{rgb}{.70,.95,1} 
\newcommand{\kms}{\mbox{km~s$^{-1}$}}
\newcommand{\x}{\mbox{$\times$}}
\newcommand{\Mo}{\mbox{M$_{\odot}$}}

\def\sbu{${\rm mag\,\,arcsec^{-2 }} $ \ }
\newcommand{\Mv}{\mbox{$M_{\mbox{\tiny V}}$}}
\newcommand{\Ha}{\mbox{H$_{\alpha}$}}
\newcommand{\Hb}{\mbox{H$_{\beta}$}}

\newcommand{\HI}{{\sc H\,i}} 
\def\HII{H\,{\small II}}
\newcommand{\OIIIb}{\mbox{[OIII]$_{\lambda 5007}$}}
\newcommand{\OIIIt}{\mbox{[OIII]$_{\lambda 4363}$}}
\newcommand{\NIIa}{\mbox{[NII]$_{\lambda 6548}$}}
\newcommand{\NIIb}{\mbox{[NII]$_{\lambda 6584}$}}
\newcommand{\SIIa}{\mbox{[SII]$_{\lambda 6717}$}}
\newcommand{\SIIb}{\mbox{[SII]$_{\lambda 6731}$}}

\title[Identification of old tidal dwarf galaxies]{Identification of old tidal dwarfs near early-type galaxies from deep imaging and \HI\ observations}
   
\author[
Pierre-Alain Duc et al.]{\parbox{\textwidth}{
Pierre-Alain Duc,$^{1}$\thanks{E-mail:\texttt{paduc@cea.fr}}
Sanjaya Paudel,$^{1}$
Richard M. McDermid,$^{2}$
Jean-Charles Cuillandre,$^{3}$
Paolo Serra,$^{4}$
Fr\'ed\'eric Bournaud,$^{1}$
Michele Cappellari,$^{5}$
Eric Emsellem$^{6,7}$
}\vspace{0.4cm}\\ 
\parbox{\textwidth}{$^{1}$Laboratoire AIM Paris-Saclay, CEA/IRFU/SAp, CNRS/INSU, Universit\'e Paris Diderot, 91191 Gif-sur-Yvette Cedex, France\\
$^{2}$Gemini Observatory, Northern Operations Centre, 670 N. A`ohoku Place, Hilo, HI 96720, USA\\
$^{3}$Canada-France-Hawaii Telescope Corporation,	65-1238 Mamalahoa Hwy., Kamuela, Hawaii 96743 USA    \\
$^{4}$CSIRO Astronomy and Space Science, Australia Telescope National Facility, PO Box 76, Epping, NSW 1710, Australia\\
$^{5}$Sub-department of Astrophysics, Department of Physics, University of Oxford, Denys Wilkinson Building, Keble Road, Oxford OX1 3RH\\
$^{6}$European Southern Observatory, Karl-Schwarzschild-Str. 2, 85748 Garching, Germany\\
$^{7}$Centre de Recherche Astrophysique de Lyon, Universit\'e Lyon 1, Observatoire de Lyon,  Ecole Normale Sup\'erieure de Lyon, CNRS, UMR 5574, 9 avenue Charles Andr\'e, F-69230 Saint-Genis Laval, France\\
}}

\begin{document}
\date{  Accepted for publication in MNRAS }

\pagerange{\pageref{firstpage}--\pageref{lastpage}} \pubyear{2013}

\maketitle

\label{firstpage}

\begin{abstract}
It has recently been proposed that the dwarf spheroidal galaxies located in  the Local Group disks of satellites (DoSs) may be tidal  dwarf galaxies (TDGs) born in a major merger at least 5 Gyr ago. Whether TDGs can live that long is  still poorly constrained by observations. 
As part of deep optical and \HI\   surveys with the CFHT MegaCam camera and Westerbork Synthesis Radio Telescope made within the ATLAS$^{\rm 3D}$ project, and follow-up spectroscopic observations  with the Gemini-North telescope, we have discovered old TDG candidates around several early-type galaxies. At least one of them has an oxygen abundance  close to solar, as expected for a tidal origin. This confirmed pre-enriched object is located within the gigantic, but very low surface brightness, tidal tail that emanates from the elliptical galaxy, NGC~5557. An age of  4 Gyr  estimated from  its  SED fitting makes it the oldest securely identified TDG ever found so far.  We  investigated the structural and gaseous properties of the TDG  and  of a  companion located in the same collisional debris, and thus most likely of tidal origin as well. Despite several Gyr of evolution close to their parent galaxies, they kept a large gas reservoir. Their  central  surface brightness is  low and their   effective radius much larger than  that of  typical dwarf  galaxies of the same mass.  This possibly provides us with criteria to identify  tidal objects  which can be more easily checked than the traditional ones  requiring deep spectroscopic observations. In view of the above, we discuss the survival time of TDGs and question the tidal origin of the DoSs.

 \end{abstract}

\begin{keywords}
galaxies: abundances --
galaxies: dwarf --
galaxies: elliptical and lenticular, cD --
galaxies: fundamental parameters --
galaxies: interactions 
\end{keywords}

\section{Introduction}

The discovery that a large fraction of Local Group (LG) dwarf galaxies are located within narrow planar  structures, the  so-called Disk of Satellites (DoSs), first speculated by \cite{Lynden-Bell76}, and confirmed through a number of  deep surveys,  raised the question of their origin  \citep{Metz07,Ibata13}.
Whether  conventional $\Lambda$CDM cosmology can produce or not such DoSs is actively debated \citep{Libeskind11,Pawlowski12,Bellazzini13}.
Alternatively, the presence  of  DoSs might simply be accounted for if all dwarfs were  formed simultaneously within a single parent structure, e.g. the collisional debris of a merger  \citep{Hammer13}.  Objects  born with that process are known  as  tidal dwarf galaxies (TDGs). 
The  importance of TDGs among the  dwarf population is rather controversial.
On the one hand,  the idealized numerical simulations of galaxy-galaxy collisions made by \cite{Bournaud06} suggest that only fine-tuned orbital parameters, mass ratio and initial gas content are needed for a merger to produce  long-lived TDGs.    \cite{Bournaud06}   estimate that, at most, ten percent of dwarfs are of tidal origin.  On the other hand, according to other numerical models \citep{Dabringhausen13} and an extrapolation at high redshift of the TDG production rate   \citep[e.g.][]{Okazaki00},   the  majority of nearby dwarfs should be of tidal origin.

The actual number of TDGs in the Universe is  poorly constrained by  observations. 
All  TDGs so far unambiguously identified and studied in detail are found in  on--going or very recent mergers: they are still linked to their parents by an umbilical cord -- the tidal tail in which they were born --, are extremely gas-rich compared to other dwarfs  and most likely have just become gravitationally bound \citep[see review by][]{Duc12}.  Among the  TDG candidates identified so far with \Ha\ or UV surveys  \citep[e.g.][]{Weilbacher00,Mendes01,Neff05,Hancock09,Smith10,Kaviraj12,Miralles-Caballero12}, a large fraction  are  certainly  gravitationally unbound star-forming knots that will quickly dissolve, or young compact star clusters that might evolve into  globular clusters. 
In such conditions, estimates of the TDG   fraction among regular galaxies substantially vary from one study to the other. This fraction  may also depend on the large-scale environment. \cite{Kaviraj12} estimated it to 6~\%  in clusters,  \cite{Sweet14}  to  16~\% in groups, 
whereas \cite{Hunsberger96} claimed that it could be as high as  50~\%  in compact groups.

 How TDGs  evolve and for how long they survive is largely uncertain.  Objects that are not kicked out from their parents are subject to dynamical friction, the gradual loss of orbital energy plunging them into the host system again,  to  destructive tidal forces generated by the deep gravitational potential of the host galaxies  \citep{Mayer01,Fleck03}, or even to the destabilizing effect of ram pressure \citep{Smith13}. 
The scenarios leading to a tidal origin for the MW and M31 satellites assume a very old merger \citep[5--9~Gyr,][]{Hammer13}. Can TDGs survive that long, and after several Gyr of evolution resemble present day Local Group  dSphs, as proposed by     \cite{Dabringhausen13}?
Besides, the morphological and color evolution of TDGs will depend on their ability to continue forming stars, and thus to retain their original large gas reservoir.  
TDGs have a star-formation efficiency which is comparable to  that of spiral galaxies \citep{Braine01}. Thus, compared to regular dwarfs, their   gas depletion time scale is rather large, and star-formation activity may in principle continue for a Hubble time.  However, the quenching mechanisms that apply to regular satellites, such as ram pressure \citep{Mayer06} should also apply to TDGs and even be reinforced \citep{Smith13}, presumably leading to a rapid reddening of their stellar populations.

The observation of several Gyr  old TDGs  is thus most needed to investigate the various scenarios proposed so far for the long-term evolution of such objects. Unfortunately when the tidal features in which they were born gradually evaporate -- in typically 2 Gyr, according to  numerical simulations  \citep{Hibbard95,Michel-Dansac10} --,  galaxies of tidal origin become more difficult to distinguish from classical ones. In principle, clear distinguishing criteria exist: a lack of dark matter \footnote{The material forming TDGs emanate from the dark matter poor disk of spiral galaxies.}, and  an unusually high metallicity for their mass  \citep[e.g. ][]{Hunter00,Duc07,Sweet14}. In practice, such characteristics need to  be checked  with high-resolution spectroscopic data, which are  expensive in terms of telescope time. 
This is in particular the case for the low-surface-brightness  dwarfs  in which star-formation has already  been quenched. 
The oldest TDGs disclosed so far based on their excess of metals and deficit of dark matter are located in the vicinity of  advanced but still relatively recent mergers: e.g., NGC~7252 \citep{Belles14}, or highly perturbed early-type galaxies, like NGC~4694 in the Virgo Cluster \citep{Duc07}.  These TDGs  may be 0.5-2 Gyr old.

In this paper  we detail the properties  of a sample of dwarf galaxies which are   satellites of   early-type galaxies  (ETGs). They were initially found during a systematic deep imaging survey  of ETGs  with the Canada-France-Hawaii Telescope made as part of the ATLAS$^{\rm 3D}$ project \citep{Cappellari11}. 
 Their location towards  faint  streams of stars, visible on CFHT MegaCam images,  or neutral hydrogen  clouds, imaged by the  Westerbork Synthesis Radio Telescope  \citep[WSRT,][]{Serra12}, also as part of  ATLAS$^{\rm 3D}$,   made them putative  TDG candidates  \citep{Duc11}. Their association with fully relaxed galaxies,  classified as lenticulars or ellipticals, rather than on-going mergers, ensured they are at least 2~Gyr old.
  The objects were followed up with the spectrograph GMOS installed on the Gemini-North telescope.  The spectroscopic observations are presented in  Section~\ref{spec}. Results on the metallicity  and light profiles of the TDG candidates are given in  Section~\ref{res}.  Finally, Section~\ref{dis} discusses the  implications of these observations on the long-term evolution of TDGs   and proposes new diagnostics of a tidal origin  that are  based on scaling relations. They are applied to the Local Group dwarfs.

\begin{figure*} 
\includegraphics[width=\textwidth]{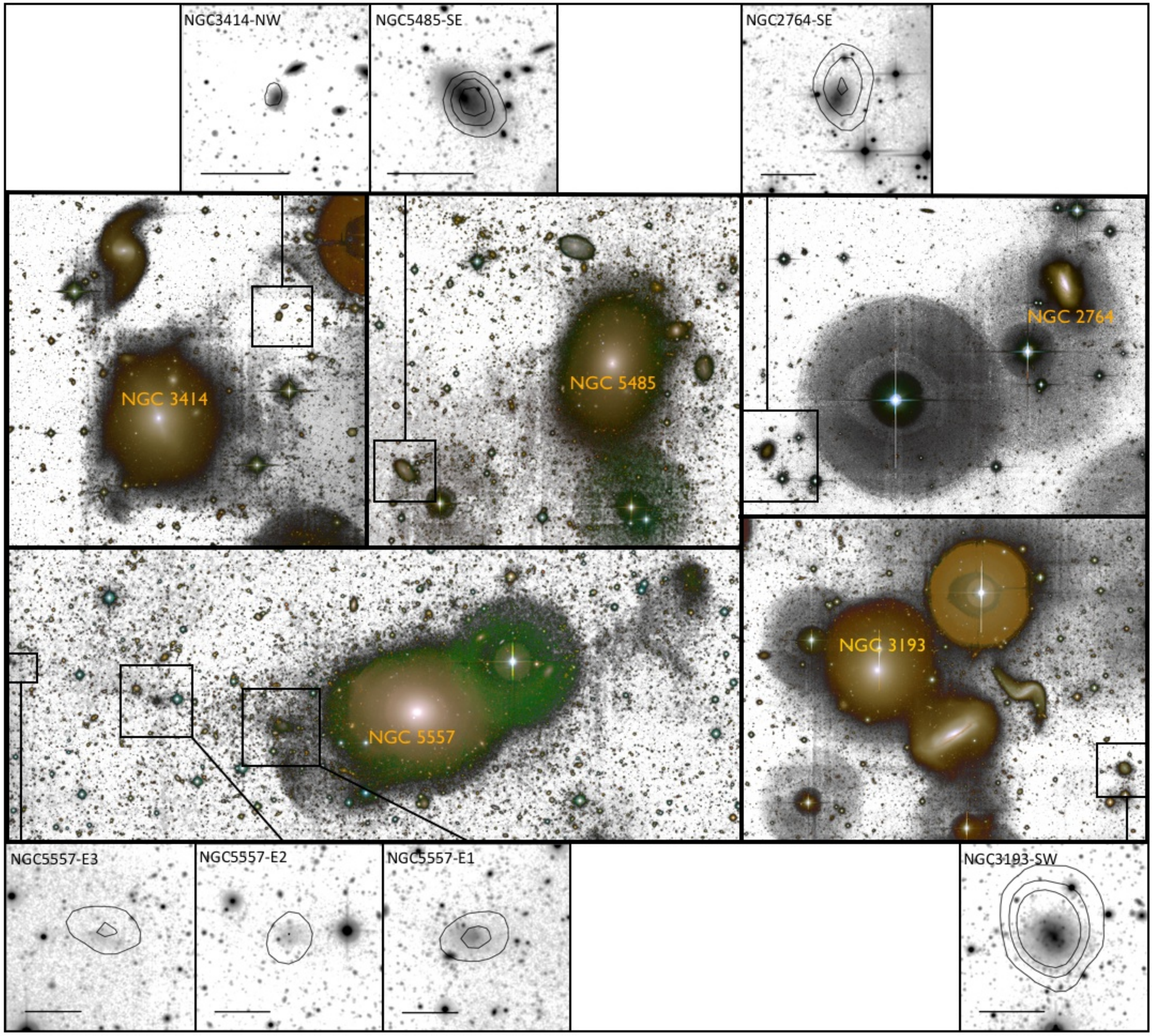}
\caption{ {\it Central panels:} composite g'+r' or g'+r'+i' MegaCam images of the  early-type galaxies hosting the dwarf galaxies  studied here.  The faintest low surface brightness features are shown as inverted grey maps for better contrast. The ETG satellites for which a  spectroscopic follow-up was carried out are indicated with the squares.   {\it Top and bottom panels:} MegaCam g'--band surface brightness maps of the pre-selected  satellites.  The field of view of each panel is 3 $\times$ 3 arcmin.  Each bar corresponds to a physical  length of 10~kpc. Greyscale levels range between 22 and 28.5 mag.arcsec$^{-2}$.   \HI\ contours from the   WSRT observations  are superimposed. Levels correspond to 0.7, 1.4 and 2.1 $\times$ 10$^{20}$ cm$^{-2}$.}
\label{ETGsample}
\end{figure*}

\section{Observations} \label{spec}

\subsection{Sample}

Fields around massive ETGs are the natural environment for the search of old TDGs as the latter form preferentially during major mergers. It is indeed difficult to extract the building material of TDGs in mergers involving galaxies with     mass ratios below 1:5  \citep{Bournaud06}. The stellar body resulting from   major merger events is most often an ETG--like object \cite[e.g.][]{Bois11}. 
So if long-lived TDGs exist, they should be present in environments which host at least one ETG.  Note however that this condition does not necessarily apply to TDGs formed in the distant Universe. Collisions at redshift of 2 or above were presumably very  different than nearby ones, involving galaxies with higher gas fraction, presumably containing massive clumps and being actively fed by external gas sources. 

The host galaxies were selected from the  ATLAS$^{\rm 3D}$ sample of nearby ($D<42$ Mpc) early-type galaxies    \citep{Cappellari11}, which benefit from a wealth of multi-wavelength data. Of relevance for this  study are the extremely deep  images obtained with the MegaCam camera installed on the CFHT \citep{Duc11} and the \HI\ maps obtained with the WSRT   \citep{Serra12}. 
The selected  ETGs  exhibit in their vicinity   low-surface brightness stellar and/or  \HI\  clouds, that might be collisional debris of past gas-rich major  mergers.  They are shown in Fig.~\ref{ETGsample}. The targets for the spectroscopic follow-up, listed in Table~\ref{loc},  are   within a distance of 200~kpc from the ETGs.  Their radio velocity, less than  600 \kms\ with respect to the hosts, ensures that they are likely  members of the ETG group. The  targets were selected in part because of their association with gas clouds; they therefore  likely contain star-forming regions and  ionized gas, allowing   us to  measure  the oxygen abundance from the nebular lines. Their surface brightness maps are shown in Fig.~\ref{ETGsample},  together with the   \HI\ 21~cm contours. 

Among the selected ETGs, NGC~5557 appears as  the most promising system for hosting old TDGs. Indeed, as shown in \cite{Duc11}, the galaxy exhibits on MegaCam images several prominent features typical of merger remnants such as 200~kpc long stellar filaments, plumes and shells.
Its eastern tidal tail hosts three blue objects, and associated with them, three isolated \HI\  clouds, referred later as NGC~5557-E1, E2 and E3.

Note that the sample is by no means complete. A systematic investigation of the origin  of  ETG satellites  is beyond the scope of this pilot study.

 \begin{table*}
 \caption{Location  of the ETG satellites}
\flushleft
\begin{tabular}{lcccccc}
\hline
Galaxy & RA &  Dec & D & $V$(\HI) & $V$(opt) & $V$(\HI)-$V$(host) \\
 & J2000 & J2000  & Mpc & \kms & \kms &  \kms \\
& (1) & (2) & (3) & (4) & (5) & (6)  \\
\hline
 NGC2764-SE   &      9:08:55.8  &   21:21:37   &     39.6 &   2680    &    2714    $\pm$12    &      -26    \\
 NGC3193-SW   &     10:17:23.3  &   21:47:58    &    33.1 &   1940   &     1964    $\pm$12   &        559  \\
 NGC3414-NW   &     10:50:49.5  &   28:03:30    &   24.5 &    1535    &    1437  $\pm$40   &         65  \\
 NGC5485-SE     &   14:08:26.8  &   54:54:32     &    25.2 &  1408     &   1389   $\pm$12    &      -519  \\
 NGC5557-E1   &    14:18:55.9   &  36:28:57     &  38.8 &    3252    &     3308  $\pm$40   &        33  \\
 NGC5557-E2     &   14:19:24.5   &  36:30:05      &   38.8 &  3196     &      --      &      -23  \\
 NGC5557-E3    &    14:19:58.1  &   36:31:53     &   38.8 &    3165      &     --       &     -54  \\
 \hline
\multicolumn{7}{@{} p{0.62\textwidth} @{}}{\small{Notes: (3) Assumed distance  (4) \HI\  heliocentric velocity determined from the WSRT moment 1 maps (5) Optical  heliocentric velocity determined from GMOS (6) Difference between the velocity of the satellite and that of the central ETG }}
 \end{tabular}
\label{loc}
\end{table*}

\begin{table*}
\caption{Structural and chemical properties  of the ETG satellites}
\flushleft
\begin{tabular}{lccccccc}
\hline
Galaxy & \Mv &$M_{*}$ & $R_{e}$ & $\mu_{g0}$ & $n$ & 12+log(O/H)  \\
  & mag & $10^{8}~\Mo$   & kpc & \sbu & dex\\
& (1) & (2) & (3) & (4) & (5) & (6)  \\
\hline
NGC~2764-SE   & -16.3  $\pm$ 0.05 & 3.9   $\pm$  1.5 & 1.4 $\pm$ 0.05 & 22.8 $\pm$ 0.1 & 0.7 & 8.3 \\
NGC~3193-SW  & -16.8 $\pm$ 0.05 & 5.6   $\pm$  2.2 & 1.3 $\pm$ 0.05 & 22.1 $\pm$ 0.1 & 1.0 & 8.3  \\
NGC~3414-NW &  -14.3 $\pm$ 0.05 & 0.4   $\pm$  0.1  & 0.8 $\pm$ 0.05  & 22.8 $\pm$ 0.1 & 0.7 & 8.2\\
NGC~5485-SE  & -17.0 $\pm$ 0.05 & 11.0   $\pm$  3.3 & 1.7 $\pm$ 0.05 & 21.7 $\pm$ 0.1 & 1.3 & 8.4 \\
NGC~5557-E1 &   -14.7 $\pm$ 0.1 & 1.2  $\pm$  0.7 (1.4)  & 2.3 $\pm$ 0.1 & 24.3 $\pm$ 0.2 & 1.3 & 8.6 \\
NGC~5557-E2  &   -13.0 $\pm$ 0.1  & 0.15  $\pm$  0.1 & 1.8 $\pm$ 0.7 & 26.1$\pm$ 0.2 & 0.6 & --   \\
\hline
\multicolumn{7}{@{} p{0.7\textwidth} @{}}{\small{Notes:  (1) Absolute magnitude in the V band determined from the MegaCam images, corrected for galactic extinction. The g' to V band conversion was done with the formula given in the SDSS DR7 on--line documentation (Lupton, 2005)   (2)  Stellar mass  estimated from the MegaCam g'  band flux and g'$-$ r' color, using the prescriptions of  \cite{Bell03}. For NGC~5557-E1, the stellar mass determined from the integrated star formation history is given in parentheses   (3) Effective radius measured from the light profile fitting in the g'--band  (4) Extrapolated central surface brightness in the g'--band (5) S\'ersic index determined for the light profile fitting  (6) Oxygen abundance estimated from the $N2$ method, and the calibration of \cite{Marino13}    }}
\end{tabular}
\label{dtb}
\end{table*}

\subsection{Observations}

\begin{figure*} 
\includegraphics[width=16cm]{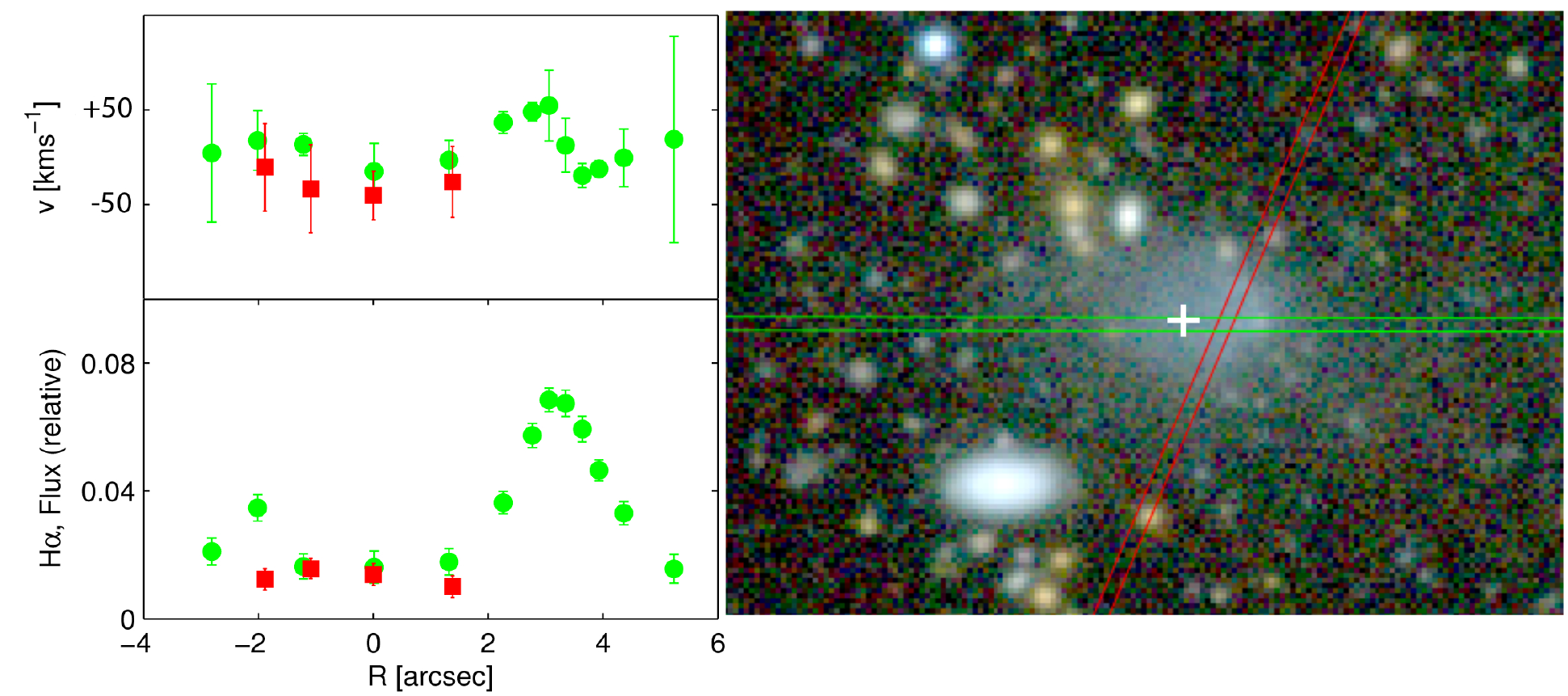}
\caption{Gas kinematic profiles of NGC~5557-E1 as extracted from the two long slits. Their directions, shown superimposed on the true color image of the dwarf to the right (composite g'+r'+i'), roughly correspond to the major and minor axis.  The colors of the data points and slits match. Position 0 on the kinematics profile indicates where the two slits intercept.  The morphological center of the dwarf, as determined from the light profile fitting, is indicated with a plus sign.
} 
\label{kinpro}
\end{figure*}

Spectroscopic observations of the TDG candidates  were carried out between March and June 2012 using Gemini Multi-Object Spectrograph (GMOS) on the 8.1 meter Gemini-North telescope (as part of program GN-2012A-Q-103). The B600\_G5307 grating was used together with a  long slit. Its width was 1.3 arcsec, leading to an instrumental resolution of 4 \AA\ FWHM. We used two slightly different grating tilts for each galaxy  in order  to fill the gap between the camera chips.  The final  wavelength coverage was   4100~\AA -- 6900~\AA. 
One single slit,  positioned  parallel to the major photometric axis,  was used for all targets, except for NGC~5557-E1 which benefited from observations along two directions, as shown in  Figure~\ref{kinpro}. Total exposure times ranged between 0.5 hour  for the most luminous dwarfs to 1.7 hour for the faintest ones, in particular those around NGC~5557.

Spectrophotometric calibration  was performed  using a single observation of the baseline standard star  (G191$-$B2B) taken separately from the science data, but with a matching instrument configuration.

\begin{figure} 
\includegraphics[width=\columnwidth]{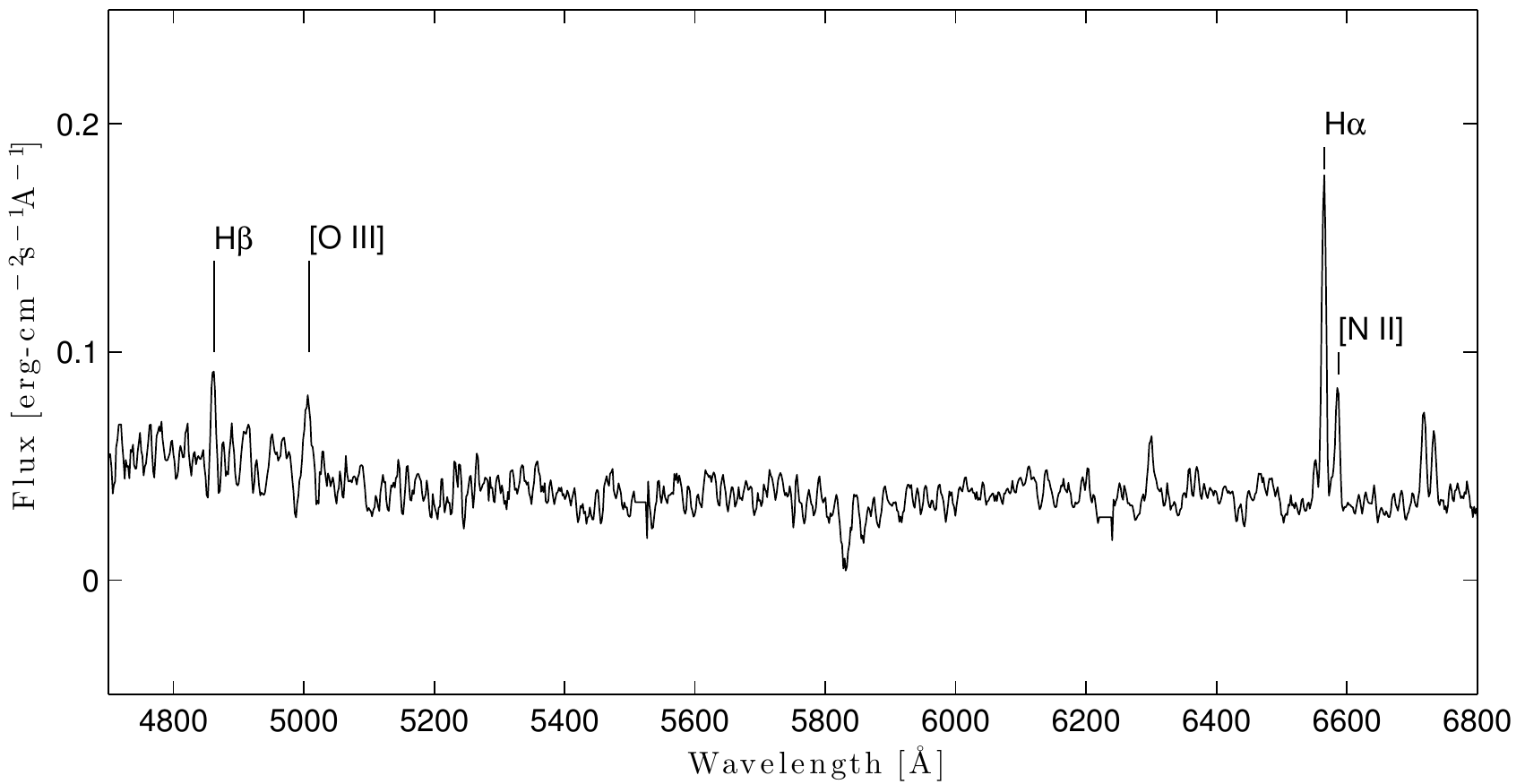}
\caption{GMOS optical spectrum of the spectroscopically confirmed TDG NGC~5557-E1.}
\label{cores}
\label{exsp}
\end{figure}

The GMOS spectra were reduced using the $gmos$ Gemini  package within IRAF. 
 Basic data reduction includes bias subtraction, cosmic ray removal, flat field correction, wavelength calibration, and flux calibration. Because some of our targets had a surface brightness much lower than the  sky brightness, special care was taken with the sky background subtraction.  
  One dimensional spectra were then extracted after rectification of the 2D  frame and summing up along the slit  in the region showing  \Ha\ emission. 
     Figure \ref{exsp} displays the  one-dimensional spectra of NGC~5557-E1, extracted and recombined from the two 2D frames acquired for this object.    The emission-line fluxes were measured from the  flux calibrated spectra using the IRAF task $splot$. Spectrophotometric data for NGC~5557-E1, our confirmed TDG as argued later,  are listed in Table~\ref{ltb}.
We did not find signatures of absorption lines  in the extracted spectra,  and thus did not correct the Balmer lines for any underlying absorption features. The  extinction coefficient was determined from the Balmer decrement assuming a theoretical value of the \Ha\//\Hb~flux ratio of 2.86 and   applied to the emission line fluxes  using the   \cite{Calzetti00} extinction law.

The photometric data   used in this paper have been measured on   CFHT / MegaCam images acquired  as part of the ATLAS$^{\rm 3D}$  survey.
 They are presented in \cite{Duc11} and Duc et al.  (2014, in prep.). Images in the g' and r'   bands are available for all our targets. NGC~5557 was also observed in the i'  band.

  \begin{table}
\caption{Spectrophotometry of NGC~5557-E1}
\begin{tabular}{lc}
\hline
   \Hb     & 100  $\pm$   18\\ 
   \OIIIb       & 92  $\pm$   18 \\
    \NIIa  &  48  $\pm$   8\\
   \Ha      & 307  $\pm$   12 \\
    \NIIb      & 124  $\pm$   8 \\
  \SIIa   & 109  $\pm$   4 \\
  \SIIb  &  82  $\pm$   6 \\
\hline
Extinction coefficient at \Ha  &  0.2 $\pm$0.1  \\
\Ha\ flux   &  6.4 $\pm$0.1   10$^{-15}$ erg s$^{-1}$ cm$^{-2}$\\
SFR(\Ha)  &  7.2 $\pm$0.1 \x  10$^{-3}$ M$_{\sun}$ yr$^{-1}$\\
\hline
\end{tabular}
\label{ltb}
\end{table}

\section{Results}\label{res}
We determined the spectrophotometric properties of all our targets, with the aim of confirming the physical association of the dwarfs with the host galaxy, and testing their tidal origin using a deviant gas-phase metallicity  as a primary criterion.

\subsection{Velocities and physical association}

We derived the radial velocity of all observed dwarf galaxies by taking the mean redshift of all detected  emission lines.  The calculated uncertainties are the  standard deviation of the measured radial line velocities. 
The optical velocities of the targets are listed in Table \ref{loc}  together with the \HI\  velocities. The difference between the optical and radio velocities ranges between 20 and 100 \kms\,  confirming the physical association between the stars and the gas for all ETG satellites.
The absence of emission lines in NGC~5557-E2 and NGC~5557-E3 (despite the presence of  \HI\ gas) did not allow us to determine their optical redshift. Furthermore due to the low surface brightness of these galaxies, no stellar continuum could be extracted.
Note however that the  \HI\  velocities of E2 and E3 also match that of E1, suggesting that they all belong to the same structure  \citep{Duc11}.

\subsection{Spectrophotometric properties}

A physical determination of the  gas-phase metallicity requires knowing  the  electron temperature, usually estimated measuring the \OIIIt\ emission line. Unfortunately, this emission line is extremely weak  and not detectable in the faint dwarf galaxies studied here.
  We have instead estimated the oxygen abundance using the so-called $N2$ empirical method,   which makes use of the strong \Ha\ and [NII]  line fluxes.
 The method, although indirect --  it assumes a standard N/O abundance ratio -- relies on the flux ratio between two lines with close wavelengths and thus has the main advantage of being little affected by uncertainties in the flux calibration or dust extinction.
Systematics errors of the method are about  0.2 dex. 
 The  oxygen abundances determined with  the recent calibration of  $N2$ by \cite{Marino13}  are listed in Table~\ref{dtb}.

\begin{figure} 
\includegraphics[width=9cm]{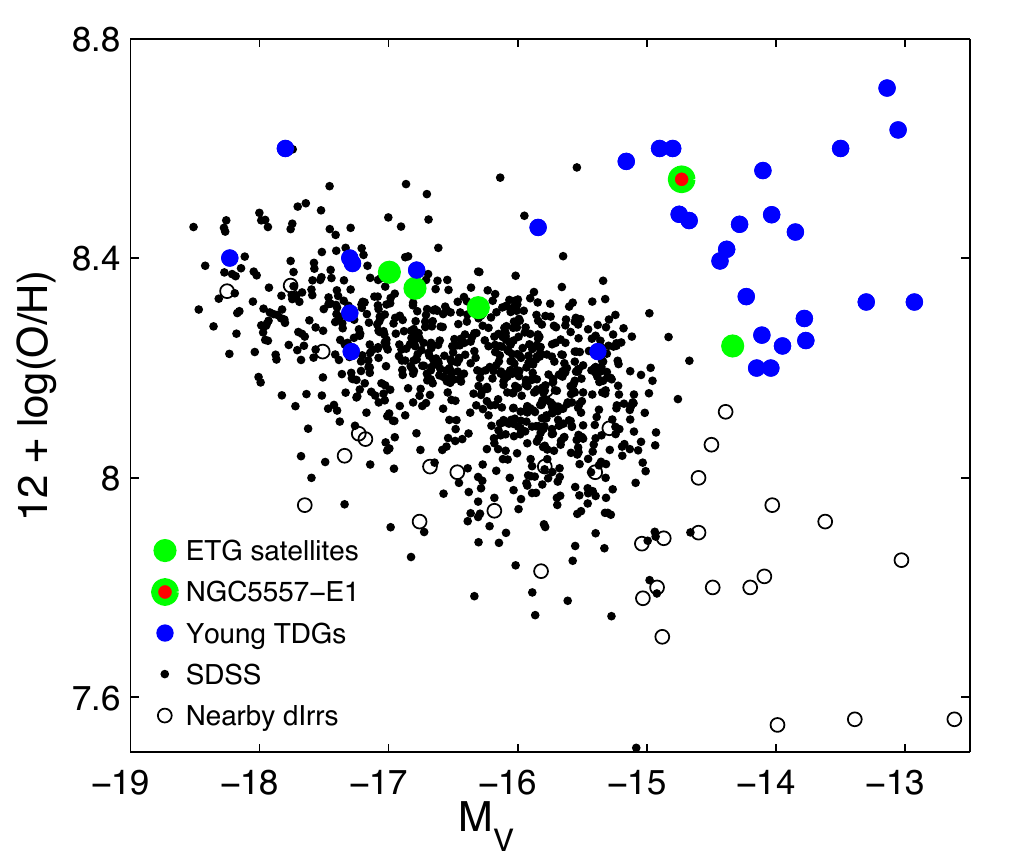}
\caption{Gas phase oxygen abundance versus absolute magnitude  for various samples of star-forming dwarf galaxies. The open circles are nearby dwarf irregular galaxies \citep{Richer95,vanZee06}; the small black dots are SDSS dwarf galaxies (see main text for details). The larger green dots are the ETG satellites  studied here, with the spectroscopically confirmed old TDG pointed with the central red dot. Young TDGs from the literature are displayed with the blue dots   \citep{Weilbacher03,Boquien10}.  Note that the oxygen abundances of the SDSS dwarfs and ETG satellites have been determined with the same $N2$ method and calibration. Those for the other dwarfs, compiled from the literature, were estimated using various  methods.}
\label{masmet}
\end{figure}

Our estimated  oxygen abundances are plotted in Fig.~\ref{masmet} against the absolute V--band magnitude.   On this figure, we added for reference  samples of young tidal dwarf galaxies from the literature  \citep{Weilbacher03,Duc07}, of  nearby  star-forming dwarf irregular galaxies from  \cite{Richer95} and \cite{vanZee06}, as well as 
 dwarf galaxies queried from the SDSS archives  having  g-r $<$ 0.4  mag, a redshift range between 0.007 and 0.015, and oxygen abundances derived by ourselves with the same $N2$ method. 
 Whereas pristine dwarfs globally follow, though with some scatter, a luminosity/mass-metallicity relation, TDGs do not  and have a too high metallicity for their luminosity. This reflects their very origin -- they are made of pre-enriched material -- and is a strong criterion to identify recycled objects. 
On this relation, the dwarfs studied here  lie between the locus of normal dwarfs and those of TDGs, leaving open the possibility that some of them may in fact be of tidal origin. 
One object suffers no ambiguity: NGC~5557-E1. With an oxygen abundance of 12+log(O/H)=8.6, i.e. close to solar \citep{Asplund05} \footnote{The empirical method based on the combination of  \OIIIb\ and \NIIb\ line -- the so called O3N2 method,  as calibrated by \cite{Marino13} -- , would give an abundance close to that obtained with $N2$:  12+log(O/H)=8.5,  but with a larger measurement uncertainty due to the faintness of the  \OIIIb\  line flux  (see Fig.~\ref{exsp}).}, the dwarf deviates by   0.6 dex from the standard luminosity-metallicity relation: the difference is much higher than the scatter of the relation and systematic uncertainties in the metallicity measurement of about 0.2~dex.

\subsection{Photometric properties}
The data reduction of the ultra--deep MegaCam images and quality control are presented in \cite{Duc11}.
The  stellar masses  of the dwarf satellites were estimated from the MegaCam g'  band flux and g'$-$ r' color, using the stellar mass to light ratios given in  \cite{Bell03}.

We used the IRAF $ellipse$ task to derive the g'--band light profile of the dwarfs. Before performing the ellipse fitting, we  subtracted the  faintest point-like sources from the images, replacing them with the surrounding sky and  masked the  background and foreground larger and brighter  objects.
In the case of NGC~5557-E1, which, at the depth of our images, is located within the extended stellar halo of its host galaxy, a model of the host was  first subtracted from the image to minimize the contamination. 
The  ellipse fitting of the dwarf was initially performed   fixing the centre  of the galaxy while  allowing a variation of the ellipticity and position angle (PA). 
To obtain the surface photometric parameters, a final run of  $ellipse$ was  done, this time fixing the PA at a value corresponding to the average of  the PAs at the outer isophotes. 

\begin{figure} 
\centerline{\includegraphics[width=0.7\columnwidth]{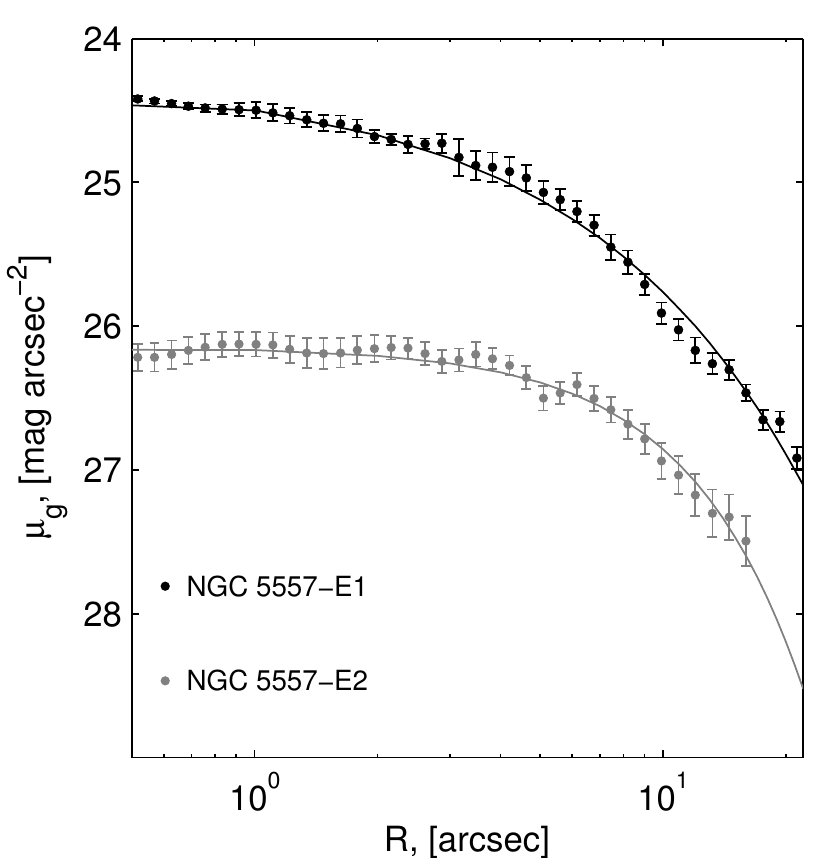}}
\caption{
g'--band surface brightness profiles along the major axis of the two confirmed old TDGs in our sample.   S\'ersic profiles are superimposed. Parameters are given in Table~\ref{dtb}.
} 
\label{lprof}
\end{figure}

\begin{figure} 
\includegraphics[width=9cm]{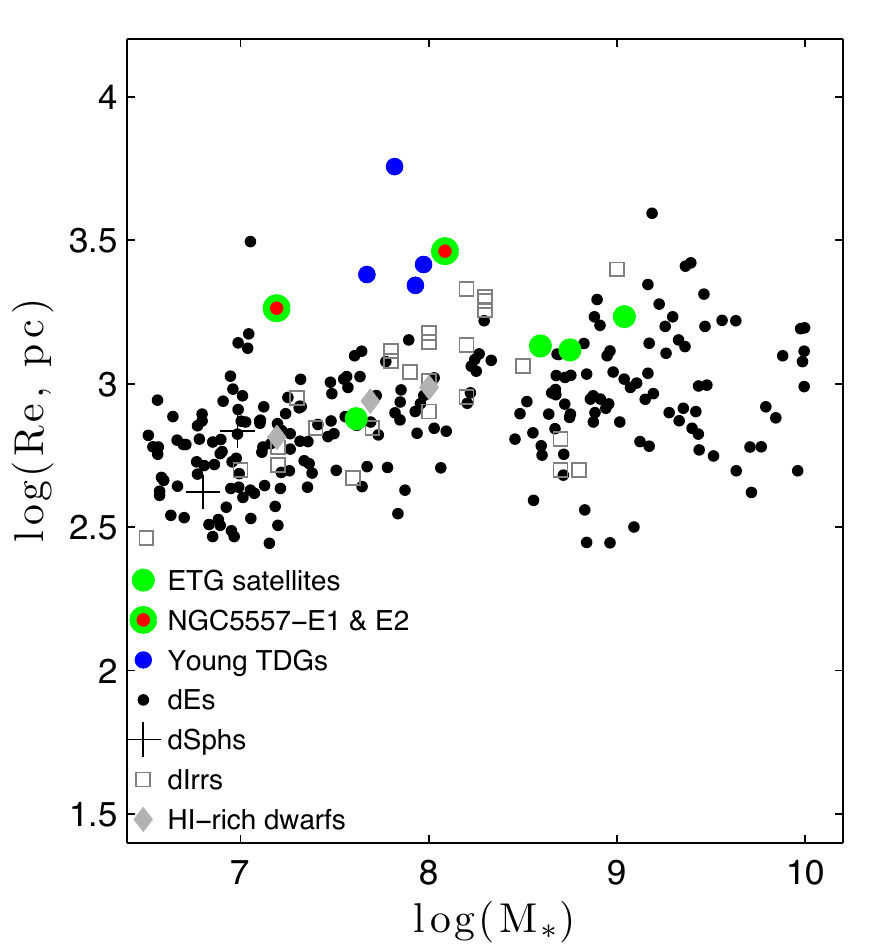}
\caption{Effective radius versus stellar mass  for various samples of dwarf galaxies. The small black points resp. crosses correspond to the compilation of dEs resp. dSphs made by Misgeld \& Hilker (2011). The grey open squares are nearby dwarf irregular galaxies extracted  from the sample of Kirby et al. (2008), and the grey diamonds are \HI\--selected star--forming dwarfs from Duc et al. (1999). The larger green dots are the ETG satellites  studied here, with the confirmed old TDGs  pointed with the central red dots. Total  stellar masses were estimated from the MegaCam g' and r' bands, using the prescriptions of  Bell et al. (2003).
Young TDGs from the literature are displayed with the blue dots. 
}
\label{sizelum}
\end{figure}

\begin{figure} 
\includegraphics[width=9cm]{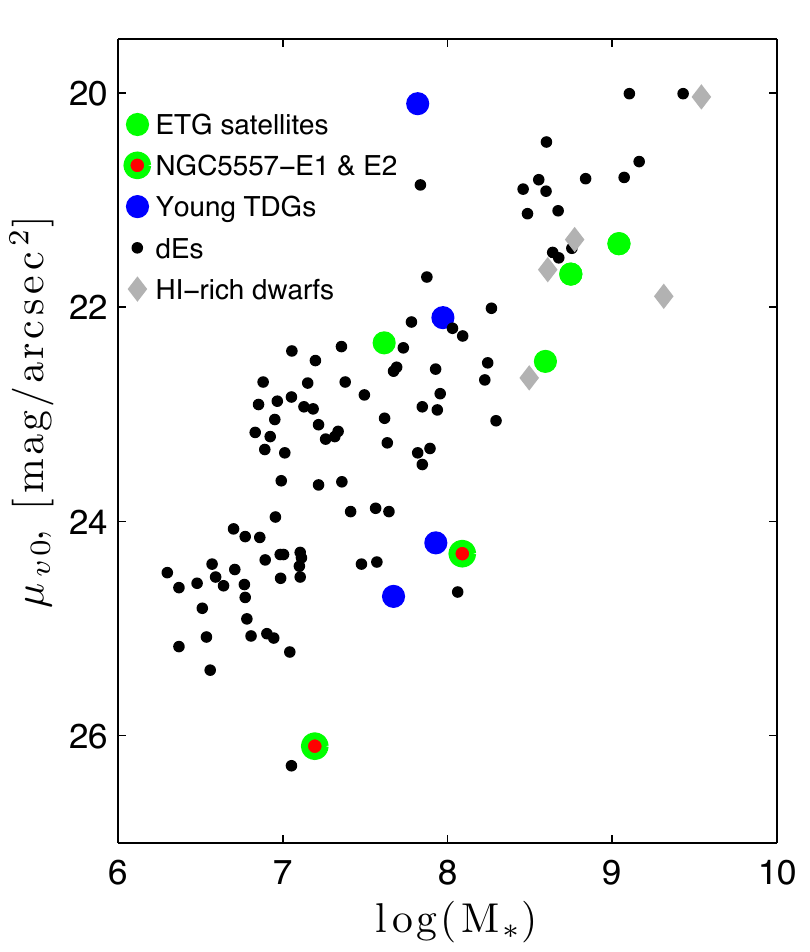}
\caption{Extrapolated central surface brightness in the V band  versus stellar mass for various samples of dwarf galaxies. The black dots  correspond to the sample of early-type dwarfs from Misgeld et al. (2008).  The grey diamonds are star-forming \HI\--selected star--forming dwarfs from Duc et al. (1999). The larger green dots are the ETG satellites  studied here, with the confirmed old TDGs pointed with the central red dots. Young TDGs from the literature are displayed with the blue dots.
} 
\label{sbf}
\end{figure}

As  examples, the  light profiles of NGC~5557-E1 and NGC~5557-E2 are shown in Fig.~\ref{lprof}. They are well fitted by  close-to-exponential  S\'ersic profiles  \citep{Sersic68}, with  an index  $n$ of resp. 1.3 and 0.6.  Due to its extremely low surface brightness and disturbed morphology, the light profile of NGC~5557-E3 could not be properly determined.
The derived  effective radii, $Re$, and  central surface brightnesses of all the dwarfs in our sample are listed in Table~\ref{dtb}, and plotted against their stellar mass in Figs~\ref{sizelum} and \ref{sbf}, together with various samples of nearby dwarf galaxies, either star-forming or passive, taken from the literature. We also  added a sample of young TDGs, for which we have measured the effective radius with the same method (see details on the data and measurements in Table~\ref{rtb}).  
 The uncertainties listed in Table~\ref{dtb} correspond to the errors in the photometric measurements (absolute magnitude, stellar masses), and galaxy profile modeling  (effective radius).
With  an effective radius as large as  resp.  2.3 and 1.8~kpc,  NGC~5557-E1/E2 occupy, like the young TDGs,  the upper locus of the size--mass relation. 
Besides, the two dwarfs also stand out in the  central surface brightness versus stellar mass relation, lying on the low-surface-brightness edge of the relation.

 \nocite{Misgeld11}
  \nocite{Kirby08}
 \nocite{Duc99}
 \nocite{Bell03}

\nocite{Gavazzi05}
\nocite{Misgeld08}

\begin{table*}
\caption{Stellar properties and surface brightness parameters of a sample of  young  TDGs}
\flushleft
\begin{tabular}{lcccccl}
\hline
Galaxy &  \Mv &$M_{*}$ & $R_{e}$ &   $\mu_{0}$  &$ n$ & Ref\\
 & mag & $10^{7}~\Mo$ &  kpc &  \sbu   & & \\
& (1) & (2) & (3) & (4) & (5)  & (6)  \\
\hline
ARP~105S & $-$16.9 & 6.6 & 5.7 & 20.1 & 1.7 & NTT/EMMI ,a, c\\
NGC~7252E & $-$14.6 & 8.5  & 2.2 & 24.2 & 0.8 & ESO2.2m/WFC, d \\
NGC~7252N & $-$14.7 & 9.4 & 2.6 & 22.1 & 2.7  & ESO2.2m/WFC, c, d \\
VCC~2062 & $-$13.2 & 4.7 & 2.4 & 24.7 & 1.0 & CFHT/MegaCam, b, c \\
\hline
\multicolumn{7}{@{} p{0.6\textwidth} @{}}{\small{
Notes:  (1) Absolute magnitude in the V band   (2) Total stellar mass derived from SED fitting   (3) Effective radius measured from the light profile fitting(4)   
Extrapolated central surface brightness  in the V band (5) S\'ersic index used for the light profile fitting (6) Reference images used for the surface brightness analysis (this paper) and bibliography: a: \cite{Duc94}, b: \cite{Duc07}, c: \cite{Boquien10},  d: \cite{Belles14}  }}
\end{tabular}
\label{rtb}
\end{table*}

\subsection{Detailled properties of NGC~5557-E1}

On all  the  diagrams  discussed above: metallicity, effective radius and central surface brightness versus luminosity/mass, NGC~5557-E1 is  located at the edge of the  distributions with respect to reference classical  dwarfs, or  even appears as an outlier.  In particular, as further argued in Sect.~\ref{old},   its high relative metallicity with respect to its luminosity  leaves  little doubts  it is a  TDG. 
In the following we focus our analysis on this specific object, and present further details on its internal properties.

\subsubsection{Stellar populations}

\begin{figure} 
\centerline{\includegraphics[width=0.8\columnwidth]{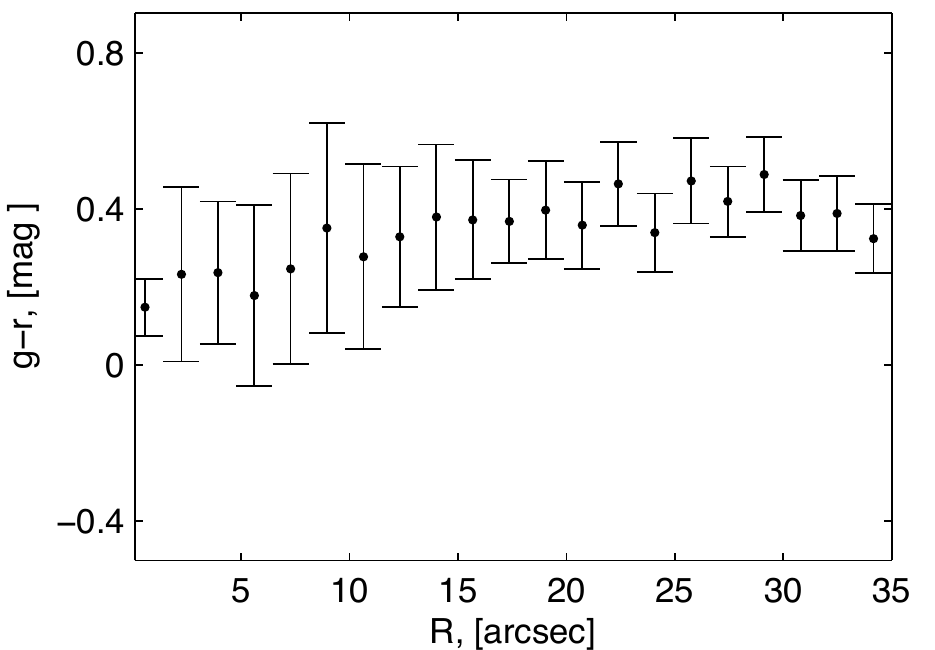}}
\caption{
g'-r' color profile of NGC~5557-E1
} 
\label{g-r}
\end{figure}

On the MegaCam composite g'+r'+i' image (see Fig.~\ref{kinpro}, right), and g'-r' color profile (see Fig.~\ref{g-r}),  NGC~5557-E1 appears as a rather blue object,  suggesting recent or ongoing star formation activity.
Indeed, the galaxy has a significant \HI\ gas reservoir:  its gas fraction, \HI\ to visible mass, is above 50\%.   Its detection in two UV-bands (FUV and NUV of GALEX) indicates ongoing star formation activity  at a rather low level: 5 \x 10$^{-3}$ M$_{\sun}$ yr$^{-1}$ \citep{Duc11}.

Our long-slit spectroscopic spectra indicate the presence of ionized gas over a range  of 5 arcsec -- about 1 kpc at the distance of the galaxy --  (see Fig.~\ref{kinpro}), confirming the presence of extended regions of star formation. The \HII\ regions are however offset with respect to the morphological center. We have obtained another estimate of the SFR from the integrated \Ha\  luminosity, corrected from dust extinction   (see Table~\ref{ltb}).
The aperture correction was performed by extrapolating the \Ha\ line measured within the long slits out to the entire galaxy. The SFR was then determined  using the calibration given by \cite{Kennicutt98}.
The  derived SFR  is within  25~\%  of that determined from the UV, confirming its rather  low value.

\begin{figure*} 
\includegraphics[width=16cm]{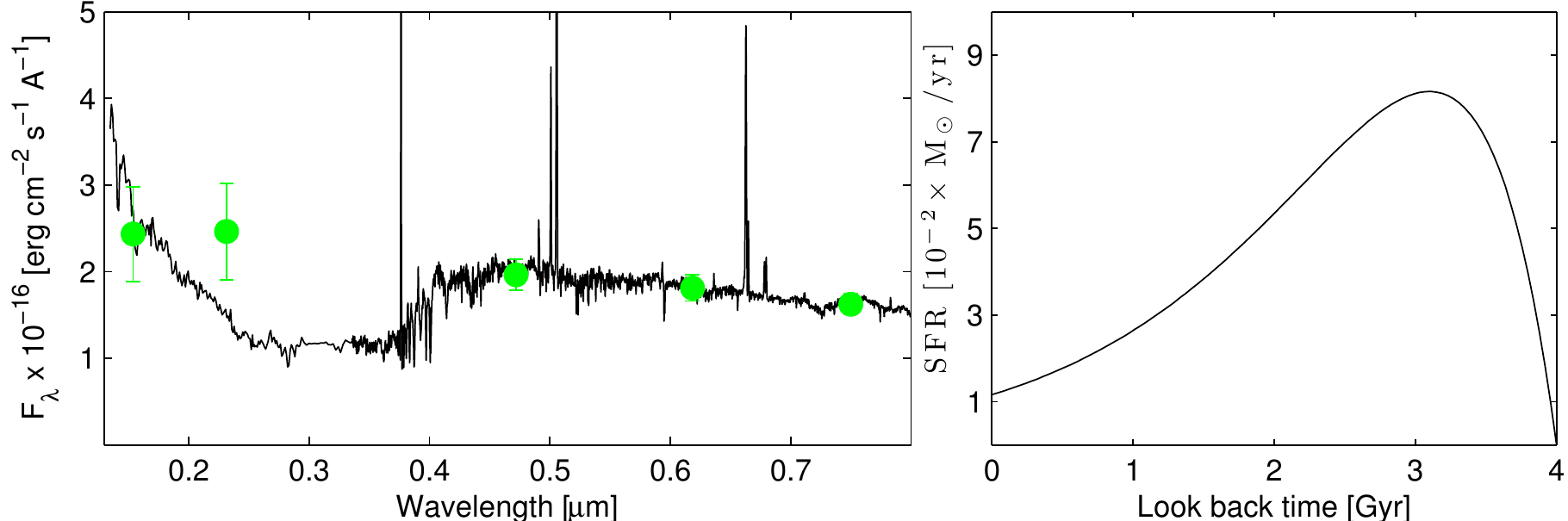}
\caption{Spectral Energy Distribution of NGC~5551-E1. An evolutionary stellar synthesis model from  Bruzual \& Charlot (2003)  is superimposed. The best fit is obtained with an exponentially declining starburst of age 4~Gyr. The corresponding star formation history is displayed to the right. } 
\label{SED}
\end{figure*}

Fitting our multi-wavelength data points  ($FUV, NUV, g', r', i'$; see values in Table~3 of \citealt{Duc11})  with the stellar population models of \cite{Bruzual03}, we made an attempt to reconstruct the star formation history of the galaxy. We constrained the model with a fixed metallicity of about solar, consistent with the measured one, varied the extinction, Av, between 0.1 and 0.3 mag, within the observed range. We further assumed  an exponentially declining star-formation history after an initial burst of star formation, with a $\tau$ value of 0.9 Gyr. 
  The best fit \footnote{which however under-predicts the NUV data point},  shown in Fig.~\ref{SED}, is obtained with  an initial burst 4 Gyr ago, and an extinction of  0.3 mag. 
  The model predicts an instantaneous current  star-formation rate of 
11$\times$10$^{-3}$ M$_{\sun}$ yr$^{-1}$, within a factor of 2 that estimated from the \Ha\ and UV. The  stellar mass derived from the integrated star formation history   is  1.4$\times$10$^{8}$ M$_{\sun}$,  a value very close to that obtained from the g' band magnitude and color, 1.2$\times$10$^{8}$ M$_{\sun}$. 
As argued in Section~\ref{old}, this basic model  gives a hint on the formation time of the object consistent with a tidal origin.

\subsubsection{Kinematics}
The kinematical profile of NGC~5557-E1 was derived from our long-slit spectroscopic data,  fitting a Gaussian to the  available emission lines, mainly  \Ha\ and [NII]. Depending on the emission line flux strength, we averaged the spectrum by 3-5 pixels in the spatial direction to increase the signal-to-noise ratio. For each position along the slits, an estimate of the mean velocity and error was determined averaging the velocity from the different emission lines. With the  grating used and seeing at time of observations, the uncertainties are estimated as approximately  $\pm$30~\kms.
 The derived velocity profiles in two directions are  shown in Fig.~\ref{kinpro}, together with the slit positions.
 The observed projected velocity gradients are about 70 \kms\ along the major axis and about 20  \kms\  along the minor one.  
  Note that this is a crude method to determine the internal kinematics of a galaxy.  First it may in principle  be  affected by slit effects \footnote{In  this particular case, the \Ha\ emission seems rather diffuse, limiting the slit effect, i.e.  the artificial velocity gradients induced by the unknown position of the emission line regions within the slit width.  The fact that the velocities at the location where the two slits intercept are within 20~\kms\ of each other is a further indication that there is no major slit effect.}. Besides, as shown in Fig.~\ref{kinpro}, the kinematical and morphological centers do not seem to coincide. 
 The  evidence for  rotation  in this galaxy,  if any,  should be confirmed by integral field  spectroscopic observations or sensitive high-resolution \HI\ observations.

\section{Discussion} \label{dis}

\subsection{The dwarfs around NGC~5557: the oldest TDGs so far identified in the nearby Universe}
\label{old}

Among the satellites of the early-type galaxies studied in this paper, one object stands out: NGC~5557-E1. With an oxygen abundance of about solar, in excess  by about   0.6 dex  with respect to the metallicity--luminosity relation, it is undoubtedly made of pre-enriched material. Such an oxygen abundance is typical of galaxies of stellar mass of about $2 \x 10^9~\Mo$ (see Fig.~\ref{masmet} and \citealt{Tremonti04}), i.e. about 10 times more massive than NGC~5557-E1. 

The MegaCam images revealed that NGC~5557-E1 is located within a giant stellar stream. The total stellar mass in these collisional debris is estimated to about 2--4\% that of NGC~5557  \citep{Duc11}, i.e. about $5 \x 10^9~\Mo$. Thus in principle there is enough material in the tails to contain the debris of a pre-existing satellite that would have lost  90\% of its initial mass. However, as seen in Fig.~\ref{ETGsample}, these debris are distributed over large distances (370~kpc), are at some locations very broad and at others have shell like structures. This makes extremely unlikely the hypothesis that they all originate in a minor merger involving a $2 \x 10^9~\Mo$ satellite, and that  NGC~5557-E1 is just its nucleus. Furthermore, such a scenario would predict a  red remnant, and an S-shape for the tidal tails on each side of the object. This is not observed. 
NGC~5557-E1 is a diffuse, blue object, located  within a long, straight tail which includes other gas condensations.

In such conditions, we can exclude that NGC~5557-E1 is the remnant of a disrupted, pre-existing satellite. This makes it a genuine tidal dwarf galaxy. We could not estimate the metallicity of its companion galaxy, NGC~5557-E2, because of its very low surface brightness and lack of ionized gas. However, given its location, further out along the same tidal tail,  its \HI\ velocity close to that of E1, it has most likely the same tidal origin. This might also be the case for NGC~5557-E3, located at the apparent tip of the tidal tail; however, its very perturbed, elongated morphology raise doubts on whether this condensation is  gravitationally bound -- another characteristic to warrant a TDG status.
 Thus, in the following for simplicity, we will call E1 and E2 confirmed old TDG candidates, although only the metallicity of E1 could be derived.

The lack of dark matter is certainly the most robust criterion to identify tidal dwarfs; it is also however  the most difficult to check. In the case of NGC~5557-E1, a velocity curve in the ionized gas  could be measured, with some evidence of a small gradient, perhaps indicating some rotation and that the galaxy is gravitationally bound. On the \HI\  WSRT map, a small velocity gradient of about  $\sim$30 kms$^{-1}$ is observed along the dwarf (see Fig. 12 in \citealt{Duc11}). Whether it traces the kinematics of the object or global streaming motions along the tail is however unclear.
Our data clearly do not have the requested velocity resolution and spatial sampling to derive the  dynamical mass and estimate the  dark matter fraction. IFU spectroscopic data would be required for that.

How old are NGC~5557-E1/E2?  The inner region of their host galaxy, NGC~5557,  is fully relaxed. It  does not contain any specific   kinematical features \citep{Krajnovic11} -- the galaxy is a slow rotator --  dust lanes, \HI\ or CO clouds, or even young stellar populations that would indicate a recent major gas-rich merger.  As shown among others by \citet{diMatteo08}, merger driven starbursts are short lived with a duration activity of a few 100 Myr. 
The  loss of memory  of  a  merger event inside NGC~5557 allows us  to estimate a minimum age for  its TDGs  of at least a couple of Gyr \citep{Duc11}.  On the other hand,  the prominent tidal tails formed during the merger event are still visible on the ultra-deep images, though their average surface brightness  is very low: around 28.5~\sbu. They furthermore still contain gas clouds. Given the relative fragility of these structures and their fading with time, they should have formed at a redshift well below 1. 
The SED fitting of NGC~5557-E1 indicates that the galaxy contains a significant fraction of stars older than a couple of Gyr. The inferred star-formation history is consistent with an initial burst occurring 4 Gyr ago, which   coincides with the dynamical time scale of its parent galaxy.
If these age estimates are correct, NGC~5557-E1, and likely its companion E2,  are the oldest confirmed TDGs so far discovered in the nearby Universe.

Simulations of galaxy mergers had predicted that under specific conditions massive tidal objects could form and survive for a few Gyr   \citep{Bournaud06}. Our observations prove that this is indeed the case.

\subsection{The morphological and stellar evolution of  tidal dwarf galaxies}

 NGC~5557-E1/E2 may be used as a laboratory and give hints on how TDGs look like after several Gyr of evolution. Previous  idealized simulations of TDG-like objects, i.e. without dark matter, had made predictions on their morphological \citep{Metz07,Dabringhausen13} and chemical evolution \citep{Recchi07}. They concluded that evolved TDGs might resemble present-day dwarf ellipticals/spheroidals (dEs/dSphs).
Such simulations  were however so far weakly constrained by  observations. 

Like typical dwarf spheroidals,  the light profile of NGC~5557-E1/E2 is indeed exponential. The surface brightness profile appears smooth, without showing prominent star-forming clumps, contrary to younger star-forming TDGs located at similar distances. The true color image of NGC~5557-E1 (see Fig.~\ref{kinpro}) and  $g'-i'$ color profile (see Fig.~\ref{g-r}) show a rather homogenous stellar distribution with perhaps a weak reddening in the outskirts  where its stars mix with that of the host tidal tail.
However,  the main distinguished features  of E1/E2 are   -- besides the relatively high metallicity (for E1)--  a low central surface brightness and large effective radius, compared to other dwarf galaxies of similar luminosity/mass and even gas content. 
As shown in Fig.~\ref{sizelum}, young TDGs observed in recent mergers already have an initially large $Re$. This seems preserved when TDGs become older, indicating that either tidal dwarfs do not manage to further condense due to internal feedback and/or are subject to some tidal   stripping in their orbit around their host. 
 \cite{Dabringhausen13} argue that the lack of dark matter in TDGs could explain their deviation from the mass-size relation.
 Stellar and gas mass loss contribute to  change the internal dynamics of the galaxy and consequently its stellar distribution: the  potential well of the system becomes shallower leading to an increase of the size.

In that respect  it is worth noting that  NGC~5557-E1/E2  kept a significant fraction of their gas reservoir. For instance, after a few  Gyr of evolution, E1  only converted half of its \HI\ gas reservoir into stars \footnote{For this estimate, we assumed that the initial stellar content was negligible, in agreement with simulations that indicate that TDGs are made from gas collapse and not instabilities in the stellar component \citep{Duc04,Wetzstein07}}. With a current SFR of 5-10 \x\ $10^{-3}$ M$_{\sun}$ yr$^{-1}$, the galaxy might continue to form stars at this rate for more than 10 Gyr. This is at odds with the observations of classical dSphs which lost all of their gas.
As a matter of fact, it is puzzling to observe that  the \HI\ reservoir  managed to remain  bound to the stellar component of the old TDGs (see Fig.~\ref{ETGsample}), despite internal feedback, tidal threshing   and even more ram pressure. The simulations of  \cite{Smith13} predict that dark matter poor galaxies subject to a wind for several Gyr should experience a strong decoupling between the gaseous and stellar components.
This may indicate that the density of the hot gas  surrounding  the parent galaxy, or the relative velocities of the TDGs  with respect to this gas,  may not be as high as those assumed  in the simulations. Recently, \cite{Ploeckinger14} developed chemodynamical models showing  the inefficiency of internal feedback and  external stripping in removing the gas of TDGs.

\subsection{The identification of old TDGs: adding an extended radius as an additional criterion? }

The spectrophotometric methods traditionally used to identify TDGs require prohibitive amounts of  telescope time even for nearby objects.
Indeed, the metallicity of TDGs -- typically half solar to solar, which is  independent of  their mass/luminosity -- become deviant only for  objects with absolute V band magnitude fainter than $-15$  (see Fig.~\ref{masmet}). 
This is unfortunately the regime for which large-scale surveys such as the SDSS  lack spectroscopic data. Furthermore, the chemical method does not apply when investigating a tidal origin for dwarfs that would have been formed in the early Universe when the metallicity of the parent galaxies was as low as todays dwarf galaxies.

A remarkable result from our study is that the two confirmed  TDGs have a large effective radius (above 1.8~kpc) for their mass, a characteristics also observed in young starbursting TDGs.
Assuming that NGC~5557-E1/E2 are representative of the TDG population, this property  may provide a very practical way to identify among  dwarf satellites those of tidal origin, even without  spectroscopy.  
Instead optical images and light profile analysis might be good enough to at least pre-select TDG candidates formed long ago.

Obviously, this criterion is not unambiguous. Indeed, at least a temporary  size expansion is also expected for tidally threshed pre-existing dwarf galaxies, as illustrated in \cite{Paudel13}. However, deep images can be obtained to disclose possibly associated tidal tails. As argued above, their shape -- linear or S-like-- can then favor one hypothesis or the other.
Besides, the very long term evolution of TDGs over a Hubble time has not been probed in our study. 
It is relevant to note that our   confirmed TDGs have a very low rate of star--formation,  the morphological appearance of  dwarf spheroidals (except their large effective radius)  but still contain gas, unlike typical fully passive dwarf satellites. However, the processes that will later trigger the star-formation quenching and stellar/gas   mass loss should rather lead to an increase rather than a shrinking of the TDG size, unless bars develop in the satellites leading to a possibly temporary concentration of the stellar component \citep{Mayer01}. 
Besides,  mergers are able to produce much more compact objects, which may result in the formation of ultra-compact dwarf galaxies  \citep{Fellhauer02,Bournaud08}. Such objects which may also be long lived  were not investigated in this paper.\\

\begin{figure*} 
\includegraphics[width=\textwidth]{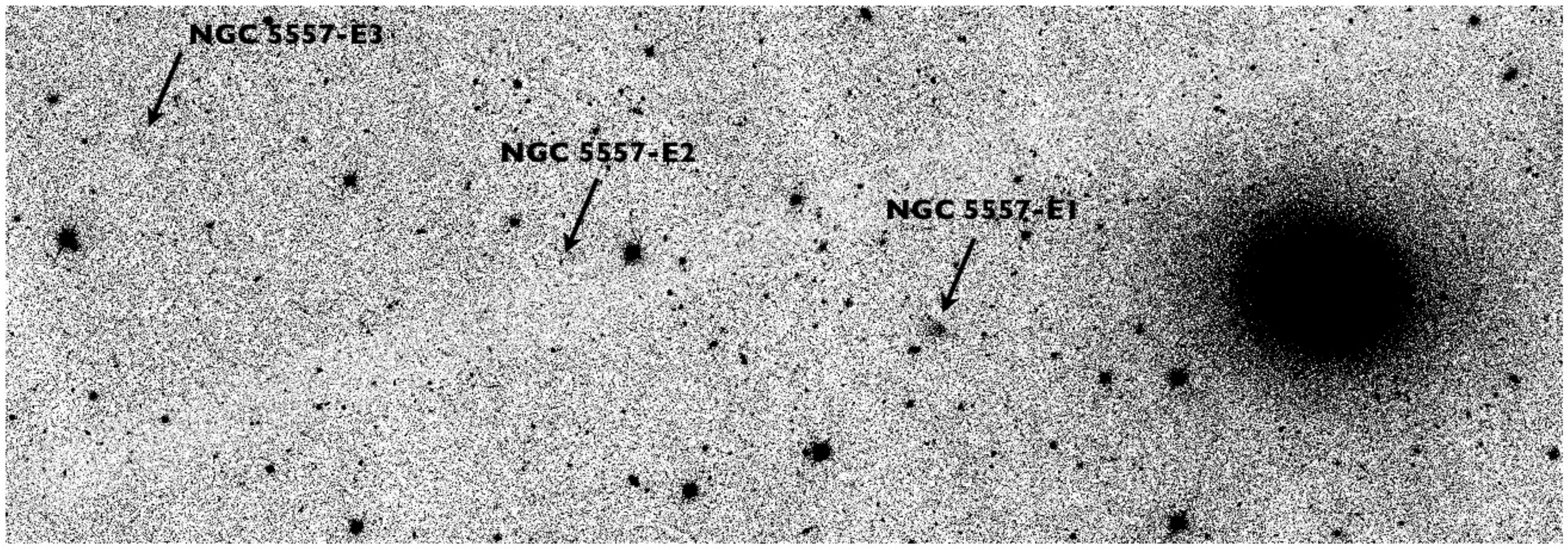}
\caption{SDSS g-band image of the field around NGC~5557
} 
\label{sdss}
\end{figure*}

 In the survey presented in this paper, besides the dwarfs around NGC~5557, 4 other satellites had been selected for a spectroscopic follow-up.  Three of them  occupy the lower envelope of the mass-central surface brightness  relation (Fig.~\ref{sbf}) and    the upper envelopes  of the mass-effective radius  (Fig.~\ref{sizelum})  and luminosity-metallicity  (Fig.~\ref{masmet})  relations. Thus, they are located towards regions where  TDGs are expected to be found.  On these diagrams, a few other objects selected from the SDSS seem to also have larger than expected effective radii and metallicities.
But none of them, including our pre-selected satellites,   deviate from these scaling relations the way the dwarfs around NGC~5557 do.
 To prove a tidal origin, further investigations should thus be carried out. Extremely  deep imaging that are able to disclose very faint low-surface brightness tidal tails linking the satellites to their parents, such as the Next Generation Virgo Cluster Survey  \citep[NGVS,][]{Ferrarese12},  or the on-going CFHT Large Programme MATLAS (Duc et al., 2014, in prep.),  might help. In the survey presented here, such direct signatures were only found for the NGC~5557 dwarfs. 
 
The large effective radius of old TDGs (for their mass) is consistent  with a low central surface brightness.
This low surface brightness means that the census of old TDGs from  imaging surveys with limited surface brightness sensitivity, such as the SDSS, may miss most of the old TDG population.  In fact NGC~5557-E1, E2 and E3 are barely visible on the Sloan images, as shown in Fig.~\ref{sdss}.

\subsection{Implications for the origin of Local Group dwarfs}
Until deep imaging/spectroscopic surveys are carried out,  the Local Group remains  the most convenient laboratory  to investigate the fraction of tidal objects among old dwarfs. If  some of the  LG dSphs are of tidal origin,  they should have been produced a long time ago,  as  the LG  does not contain any  vestige  of a recent major merger. The metallicity method, and,  more generally,   criteria relying on the chemistry   or stellar properties thus do not directly apply to these objects. However, as argued earlier,  a relatively large effective radius might be a durable characteristic of a tidal origin.

In that respect, the dwarfs located within the disks of satellites, that were claimed as being old TDGs, are not located in specific regions  of the scaling relations: only   10\% of the M31 dSphs have effective radii larger  than 1~kpc and none for the MW dSphs \citep{Brasseur11}. This  may weaken the TDG hypothesis, which was already  challenged by  the   evidence that the MW and M31 satellites are dark matter rich (assuming conventional CDM cosmology). 
Another possible difference between the  TDG discovered in this study and local dwarf satellites is their gas content. Despite several Gyr of evolution, the TDGs around NGC~5557 are still gas--rich. This reflects their origin as tidally expelled  condensations made almost entirely of gas and  suggests  a  rather low efficiency at converting gas into stars after the initial collapse  and inefficient  gas blow-away and stripping mechanisms. Whether the puzzling ability of keeping gas is something specific to the system studied here or an intrinsic characteristics of TDGs born in \HI\ tidal tails remains to be determined.

In any case, the origin of the DoSs remains mysterious: the dwarf galaxy accretion at the heart of the hierarchical cosmological models cannot easily reproduce such narrow distribution of satellites.   
One possibility to investigate  is whether  the DoSs are made of objects with an origin similar to TDGs, i.e. born within larger structures, but not strictly speaking within  tidal debris produced by major mergers, as those so far considered for the tidal dwarfs. These could be dwarfs originating in the massive clumps observed in distant gas--rich galaxies   \citep[e.g.][]{Elmegreen07,Bournaud07c} and  then kicked out  due to  clump-clump interactions or dwarfs  born together within gaseous cosmological filaments.

\section{Summary and conclusions}

We have obtained with Gemini-North optical long-slit spectra of a sample of 7 dwarf galaxies in the vicinity of  5 early-type galaxies (ETGs) selected from the ATLAS$^{\rm 3D}$ survey.  Deep optical CFHT/MegaCam images and WSRT radio observations   had previously detected around the selected ETGs   gaseous and/or stellar streams  that made the host galaxies  likely old gas-rich mergers, and their dwarf satellites  excellent tidal dwarf galaxy (TDGs) candidates. The spectroscopic observations were aimed at measuring the oxygen abundance of the ETG satellites, and investigate any deviation from the mass-metallicity relation that may  indicate a tidal origin.
We furthermore used the MegaCam images to determine the morphological properties of the dwarfs and their behavior  with respect to  standard scaling relations. We obtained the following results:

\begin{itemize}
\item 
One dwarf, referred as NGC~5557-E1,  located in the vicinity of the massive elliptical, NGC~5557,  has a gas-phase metallicity of about solar; its oxygen abundance is 0.6 dex above that typical of regular dwarf galaxies, or conversely its abundance corresponds to that of a galaxy ten times more massive.
This dwarf, together with two fainter companion objects, E2 and E3 (for which no spectra could be extracted),  lie along an extended, very low surface brightness tidal tail, most likely formed during a major merger  that occurred at least 2 Gyr ago. A burst model fitting the  spectral energy distribution of NGC~5557-E1 provides an age for the  dwarf of   about 4 Gyr, consistent with the age estimate of the merger at the origin of NGC~5557. Object E1, plus most likely its  companions located in the same tidal structure, in particular the apparently relaxed object E2,  would thus be the oldest confirmed  tidal dwarf galaxies so far identified.
\item
Like regular dwarf ellipticals/spheroidals,  NGC~5557-E1/E2 are well fitted by exponential light profiles. However, they have an unusually large effective radius, and low central surface brightness for their mass. This may be an intrinsic characteristic of old TDGs.  Besides, contrary to typical satellites  of massive galaxies, NGC~5557-E1/E2 managed to retain a large gas reservoir despite several Gyr of evolution and interaction with their parent galaxy.   They may preserve it for several additional Gyr, given their very low current star-formation rate, as measured from the UV and/or \Ha\ luminosity, unless stripping mechanisms become important. 
\item
 Among the four other ETG satellites in our survey, three that are more massive  than NGC~5557-E1/E2  tend to also have rather  low central surface  brightness and large  effective radius. However unlike the confirmed TDGs, they are not clear outliers of the luminosity-metallicity  relation. In fact, the method to identify TDGs using deviations from scaling relations  only works well for  low-mass objects. Future systematic surveys of tidal dwarfs should thus rather focus on the least massive ones, identified in deep images.
They may tell whether the large effective radius found for the old TDGs NGC~5557-E1/E2, but also for young TDGs, is a durable, intrinsic characteristic of a tidal origin. If this is the case, it is then unlikely that the dwarfs in the disk of satellites around our Milky Way and Andromeda are old TDGs made in gas--rich major mergers.

\end{itemize}

\section*{Acknowledgments}
We first express our gratitude to the anonymous referee for his very careful reading of the paper and for spotting a few inconsistencies in some  tables which forced us to check and revise  our measurements. 
This work would not have been possible without the wealth of data acquired as part of the ATLAS$^{3D}$ collaboration. We are grateful to all  team members. We warmly thank Raphael Gobat for his help with the SED fitting. 
This paper made used of spectra obtained at the Gemini Observatory, which is operated by the 
    Association of Universities for Research in Astronomy, Inc., under a cooperative agreement 
    with the NSF on behalf of the Gemini partnership: the National Science Foundation 
    (United States), the National Research Council (Canada), CONICYT (Chile), the Australian 
    Research Council (Australia), Minist\'{e}rio da Ci\^{e}ncia, Tecnologia e Inova\c{c}\~{a}o 
    (Brazil) and Ministerio de Ciencia, Tecnolog\'{i}a e Innovaci\'{o}n Productiva (Argentina).
This study was also  partly based on observations obtained with MegaPrime/MegaCam, a joint project of CFHT and CEA/IRFU, at the Canada-France-Hawaii Telescope (CFHT) which is operated by the National Research Council (NRC) of Canada, the Institut National des Science de l'Univers of the Centre National de la Recherche Scientifique (CNRS) of France, and the University of Hawaii. 
SDSS data were queried from the SDSS archives.
Funding for the SDSS and SDSS-II has been provided by the Alfred P. Sloan Foundation, the Participating Institutions, the National Science Foundation, the U.S. Department of Energy, the National Aeronautics and Space Administration, the Japanese Monbukagakusho, the Max Planck Society, and the Higher Education Funding Council for England. The SDSS Web Site is http:\/\/www.sdss.org.
The SDSS is managed by the Astrophysical Research Consortium for the Participating Institutions. The Participating Institutions are the American Museum of Natural History, Astrophysical Institute Potsdam, University of Basel, University of Cambridge, Case Western Reserve University, University of Chicago, Drexel University, Fermilab, the Institute for Advanced Study, the Japan Participation Group, Johns Hopkins University, the Joint Institute for Nuclear Astrophysics, the Kavli Institute for Particle Astrophysics and Cosmology, the Korean Scientist Group, the Chinese Academy of Sciences (LAMOST), Los Alamos National Laboratory, the Max-Planck-Institute for Astronomy (MPIA), the Max-Planck-Institute for Astrophysics (MPA), New Mexico State University, Ohio State University, University of Pittsburgh, University of Portsmouth, Princeton University, the United States Naval Observatory, and the University of Washington.  This work is supported  by the French Agence Nationale de la Recherche (ANR) Grant Programme Blanc VIRAGE (ANR10-BLANC-0506-01).  MC acknowledges support from a Royal Society University Research Fellowship.
This work was supported by the rolling grants Astrophysics at Oxford PP/E001114/1 and ST/H002456/1 and visitors grants PPA/V/S/2002/00553, PP/E001564/1 and ST/H504862/1 from the UK Research Councils. RMcD is supported by the Gemini Observatory, which is operated by the Association of Universities for Research in Astronomy, Inc., on behalf of the international Gemini partnership of Argentina, Australia, Brazil, Canada, Chile, the United Kingdom, and the United States of America. 
\label{lastpage}

\bibliographystyle{mn2e}
\bibliography{tdg-duc}

\end{document}